\documentclass[twocolumn,preprintnumbers,aps,pra]{revtex4}
\usepackage{graphicx}

\begin{document}

\title{Low-cost nanosecond electronic coincidence detector}

\author{Taehyun Kim\footnote{E-mail address: thkim@mit.edu}, Marco Fiorentino, Pavel V. Gorelik and Franco N. C. Wong}

\affiliation{Research Laboratory of Electronics, Massachusetts Institute of 
Technology, Cambridge, Massachusetts 02139}

\begin{abstract}
We present a simple and low-cost implementation of a fast electronic coincidence 
detector based on PECL logic with a TTL-compatible interface.  The detector has 
negligible dead time and the coincidence window is adjustable with a minimum 
width of 1 ns.  Coincidence measurements of two independent sources of 
Bose-Einstein distributed photocounts are presented using different coincidence 
window widths.
\end{abstract}
\maketitle

\section{INTRODUCTION}
Detecting coincidence of two events is crucial in many 
areas of science, such as quantum optics \cite{TAC, DOWN, PAIR}, 
biophysics \cite{PET}, and nuclear physics \cite{GAMMA, FUEL}.  Most 
commercially available electronic coincidence detectors are based on 
time-to-amplitude converters \cite{ORTEC} and even though they have good time 
resolution ($\sim$10 ps), they are generally expensive and have relatively long 
dead times (1 $\mu$s). Other types of commercial coincidence detectors 
\cite{SRS}, which use AND gates,
 have $\sim$5--10 ns coincidence windows and they can still be expensive. 
However, many applications require much narrower window to improve the 
signal-to-noise ratio or to match the temporal width of the triggering event.  
For example, typical Si single-photon counters for visible/near-infrared light 
have output pulses with rise times that are less than 0.5 ns, which set the 
lower limit of the coincidence window.  Moreover, in our area of interest, 
quantum optical 
information technology, multiple coincidence measurements are often needed so 
that cost and operational complexity are important considerations.

In addition to time-to-amplitude conversion, there are alternative methods to 
high-speed coincidence detection.  Measurements using an ultrafast digital 
oscilloscope on two or more channels with deep storage memory allow a long time 
history to be recorded, whose length depends on the depth of the storage memory.  
This is particularly useful for system testing and debugging purposes.  However, 
a digital storage scope is not suitable for efficient and nearly real-time 
measurements because of the need for extensive postdetection data processing.  
Personal computer-based picosecond time-resolved 
multichannel coincidence detection relies on measuring the time difference 
between two events and very high time resolution of a few picoseconds can be 
achieved, and is often used in pump-probe fluorescence lifetime spectroscopy.  
The multichannel capability and ps time resolution requires expensive hardware 
and the method is ill suited for a simple coincidence detection of two events 
within a fixed time window.  In this work we present a simple and low-cost 
circuit-board implementation of two-event coincidence detection with a variable 
coincidence time window as short as 1 ns and a small propagation delay of 
$\sim$5 ns.   The coincidence detector is based on positive emitter-coupled 
logic (PECL) with transistor-transistor logic (TTL) input/output for convenient 
interface with other electronic equipment, such as a single-photon counter.  It 
can be used in a non-triggered continuous-wave (cw) mode or in a triggered 
configuration with sub-ns gating \cite{PAIR} that can further improve the signal 
to noise ratio.  In addition, the circuitry has essentially no dead time and 
hence it can be used at rates up to hundreds of MHz.  The cost benefit is 
significant: our coincidence detector costs less than \$100\,USD. 

\section{CIRCUIT IMPLEMENTATION}
The easiest way to detect a coincidence of two events is to apply an 
AND-logic operation to the pulses from the two event detectors,
but, in general, the output pulse widths of the detectors are much longer than
the desired coincidence window of a few ns.  For example, in our photon counting 
work \cite{DOWN}, the output pulse width from a commercial Si single-photon 
counting module (SPCM) \cite{Perkin} is $\sim$30 ns.  Yet, in order to suppress 
accidental coincidences it is desirable to detect the coincidence of the two 
photons within $\sim$1 ns, limited by the $\sim$350-ps jitter of the SPCM output 
pulse.  Therefore, each pulse width should be reduced to $\sim$1 ns before the 
AND operation is applied and the AND gate should also be fast enough to detect 
the sub-ns pulse overlap.  Unfortunately TTL components, which are commonly used 
in electronic equipment including our SPCMs, are not fast enough to handle 
ns signals.  Our solution to this problem is to use PECL for the core of 
the coincidence detection system and to employ TTL--PECL translators for 
interface compatibility.  The maximum rise or fall time for PECL is $\sim$400 ps 
and hence PECL is sufficiently fast for use with the Si SPCM\@.  The translation 
from TTL to PECL does not cause a loss of timing information, but adds a
common propagation delay of a maximum of 600 ps to both channels.  The 
translation from PECL to TTL has a propagation delay of 1.5--5 ns and a rise 
time of 0.3--1.6 ns, but these do not affect the performance because this on/off 
output is fed to a counter.

\begin{figure}[ht]
  \begin{center}
    \includegraphics[width=8.5cm]{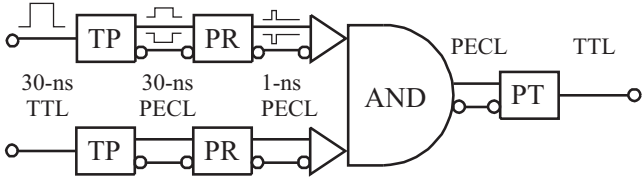}
  \end{center}
  \caption{Block diagram of coincidence detection circuit. TP: TTL-to-PECL 
translator, PR: pulse reshaper, PT: PECL-to-TTL translator.  Differential 
interconnections are used bewteen TP and PR and for the inputs and output of the 
AND gate for improved noise immunity.}
  \label{FIGURE_CIRCUIT}
\end{figure}

Fig.~\ref{FIGURE_CIRCUIT} shows the block diagram of the full circuit.
For each of the two input channels, a TTL signal derived from an event detector 
such as a SPCM is translated to PECL level (Fairchild 100ELT22MX) at the 
TTL-to-PECL (TP) block.  The resultant PECL pulse, which is as long as the input 
TTL pulse (we shall use the Si SPCM $\sim$30-ns output as the example throughout 
this work), is shortened by the pulse reshaper (PR), and the two narrowed pulses 
from the two channels go through the PECL AND gate (On Semiconductor MC10E404) 
for coincidence detection.  Finally this PECL output is translated back to TTL 
level (ON Semiconductor MC10H350) at the PECL-to-TTL (PT) block. 

A disadvantage of using PECL logic to achieve low noise and high speed 
core logic operation is that its signal routing is more complicated than other 
logic families.  Differential interconnection between components and several 
termination resistors at the output of each component are required.  The circuit 
board for PECL components requires at least three layers to provide 
a noise-free +5\,V plane and a ground plane for the TTL components.  In our 
implementation, we used a four-layer standard FR-4 laminate printed circuit 
board (PCB).  The line widths on the PCB were 0.25 mm (characteristic impedance 
$\sim$65 $\Omega$), except for the paths connecting to the BNC connectors where 
0.5 mm line widths were used to provide  50 $\Omega$ impedance matching.  These
impedances were calculated based on the 0.3-mm gap between the ground layer and 
the top layer and a dielectric constant of 4.6 for the substrate.

\begin{figure}[ht]
  \begin{center}
    \includegraphics[width=8.5cm]{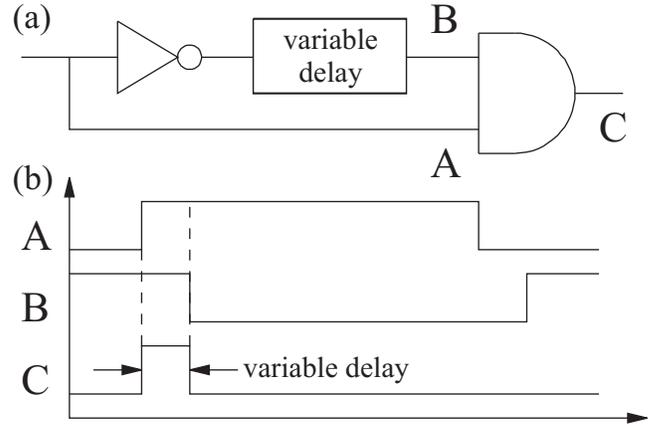}
  \end{center}
  \caption{Pulse reshaper. (a) Circuit schematic for reshaping the long input 
pulse into a short pulse. The time delay is adjusted by varying the length of 
the coaxial cable. (b) The timing of the pulses at points $A$, $B$, and $C$.}
  \label{FIGURE_SHORT_PULSE}
\end{figure}

\begin{figure}[ht]
  \begin{center}
    \includegraphics[width=8.5cm]{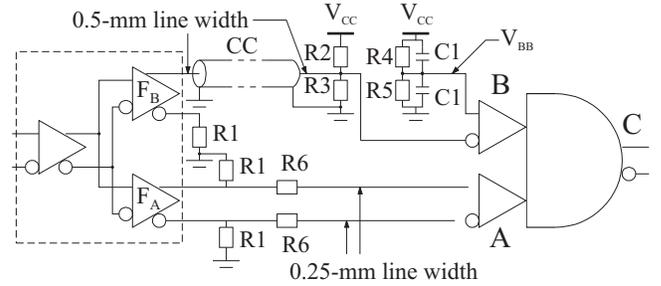}
  \end{center}
  \caption{Detailed circuit diagram of pulse reshaper. R1: 287 $\Omega$, R2: 
82.5 $\Omega$, R3: 124 $\Omega$, R4: 95.3 $\Omega$, 
R5: 261 $\Omega$, R6: 62 $\Omega$, C1: 0.1 $\mu$F, CC: coaxial cable (RG-58, 50 
$\Omega$), V$_{\rm CC}$= 5\,V, V$_{\rm BB}\approx 3.66$\,V.}
  \label{FIGURE_PR_CIRCUIT}
\end{figure}

To reshape the long input pulse into a short one, we used a NOT gate, an AND 
gate and a variable delay line as shown in the schematic of 
Fig.~\ref{FIGURE_SHORT_PULSE}.  The delay line was simply a short RG-58 coaxial 
cable whose length could be changed to determine the reshaped pulse 
duration (5 ps/mm).  The details of the pulse reshaper circuit are shown in 
Fig.~\ref{FIGURE_PR_CIRCUIT}.  The dashed box is a PECL fanout buffer (Fairchild 
100EL11M) that converts the differential input into two differential 
outputs. The buffer $F_A$ was connected to the differential input A of the AND 
gate with 62-$\Omega$ (R6) series terminations to  match the 
characteristic impedance of the line on the PCB \cite{MECL}.

The non-inverting terminal of the fanout buffer $F_B$ 
was connected to the inverting terminal of the input B of the AND gate to
implement the NOT gate. This NOT path includes the coaxial cable (CC in 
Fig.~\ref{FIGURE_PR_CIRCUIT}) to generate the time delay.  Unlike the rest of 
the circuit with short distances, this long delay line required a different 
method for impedence matching, and we chose the Thevenin-equivalent parallel 
termination scheme \cite{PECL} with a single-ended interconnection.   We did not 
use a differential interconnection to avoid the use of two length-matched 
coaxial cables for the inverting and non-inverting paths.  The non-inverting 
terminal of the AND gate input B is biased at a reference voltage of 
3.66 V, which is the mid-point between PECL logic 1 and 0, and the voltage 
difference between this reference and the signal transmitted by the coaxial 
cable serves as the differential input of the AND gate.

\begin{figure}[ht]
  \begin{center}
    \includegraphics[width=8.5cm]{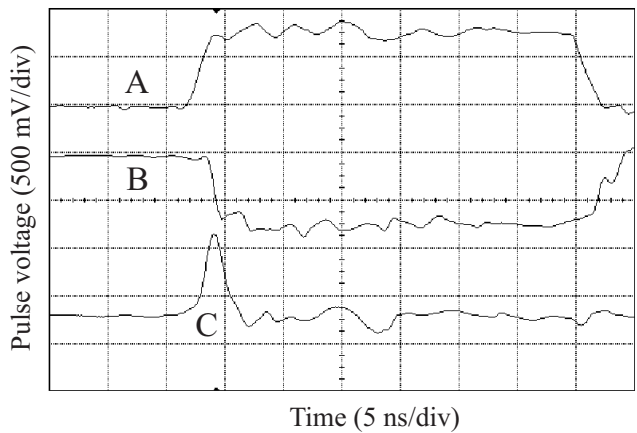}
  \end{center}
  \caption{Oscilloscope image of signals inside the pulse reshaper with a 20-cm 
coaxial delay line. The pulse shapes are limited by the analog bandwidth
of the 500 MHz oscilloscope probe ($2\times 10^9$ samples/s).}
  \label{FIGURE_PULSE_GRAPH}
\end{figure}

Fig.~\ref{FIGURE_PULSE_GRAPH} shows the two input signals and the output of the 
AND gate of the pulse reshaper, corresponding to the schematic sketches of 
Fig.~\ref{FIGURE_SHORT_PULSE}(b).  Curve (A) of Fig.~\ref{FIGURE_PULSE_GRAPH}
shows an image of the input pulse to the 
reshaper with a pulse duration of 33 
ns, while curve (B) reproduces its inverted image with a delay of $\sim$1 ns for 
the 20-cm-long coaxial delay line.  Curve (C) shows the shortened pulse of the 
pulse reshaper output with a duration of the desired width of $\sim$1 ns.

\section{TIMING CALIBRATION}

We calibrated the timing characteristics of the coincidence detector by using 
two identical pulses with a variable time delay as inputs.  The two input pulses 
were derived from the output pulse of a Si SPCM by splitting it into two signal 
pulses.  The variable time delay between $-2000$ ps and $+2000$ ps was obtained 
by using cables of different lengths in steps of less than 1 cm (50 ps). We 
measured the ratio of coincidences to singles with an 80-MHz electronic counter 
(National Instruments 6602).  This counter, however, cannot detect pulses less 
than $\sim$3.5 ns in width. Since the AND gate output pulse width is
determined by the delay lines, a timing overhead must be added.
The amount of timing overhead was determined experimentally by using 
different lengths of delay lines and measuring the effective coincidence window 
as described above.  In our circuitry, we increased the delay lines to 50 cm of 
coaxial cables (instead of 20 cm) to achieve a coincidence window of $\sim$1 ns.  
It should be clear that for different counters, the cable lengths must be 
adjusted and the coincidence windows recalibrated.  We note that care was taken 
to reduce signal distortion between the coincidence detector and the electronic 
counter by using twisted-pair connection (AWG 22 hook-up wire, $\sim$1 turns/cm) 
with series termination of 56 $\Omega$.  

\begin{figure}[ht]
  \begin{center}
    \includegraphics[width=8.5cm]{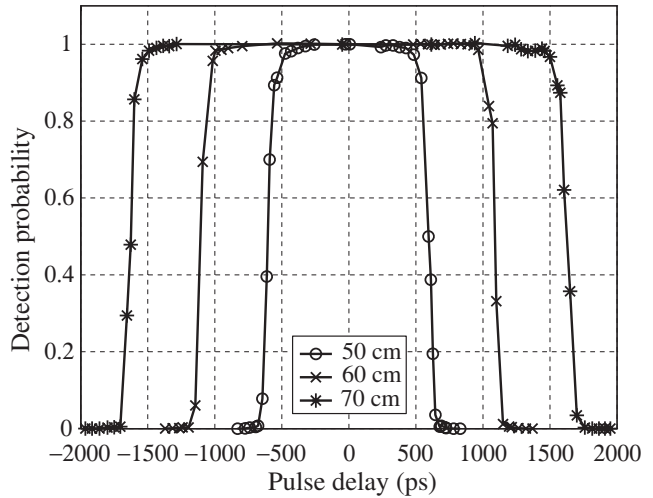}
  \end{center}
  \caption{Detection probability of coincidences vs delay between two pulses for 
different lengths of delay lines.}
  \label{FIGURE_PROBABILITY}
\end{figure}

Fig.~\ref{FIGURE_PROBABILITY} shows the timing results for the cases of 50 cm,
60 cm, and 70 cm delay lines, by plotting the coincidence probability as a 
function of the relative delay between the two input pulses.  We note that the 
width of the transitional region (defined as the region with 10$\rightarrow$90\% 
probability) is $\sim$100 ps on each side for all three lengths of the delay 
lines.  We believe that the width of the transitional region is due to the 
jitter of the triggering threshold of the counter.  To confirm this, pulses from 
a digital delay/pulse generator (Stanford Research Systems DG535, 5-ps delay 
resolution) were counted as the pulse width was varied about the minimum width 
of $\sim$3.5 ns required by the counter, and we obtained the same 100-ps 
transitional regions.  The effective coincidence window size, defined by the 
full width at half maxium (FWHM) in Fig.~\ref{FIGURE_PROBABILITY}, 
is 1.2, 2.2, and 3.26 ns for the 
50-cm, 60-cm, and 70-cm delay lines, respectively.

\section{COINCIDENCE COUNTING PERFORMANCE}

\begin{figure}[ht]
  \begin{center}
    \includegraphics[width=8.5cm]{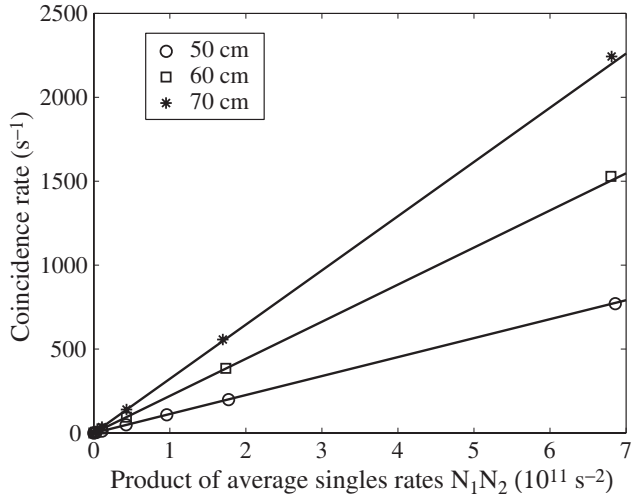}
  \end{center}
  \caption{Accidental coincidence measurements of two independent sources of  
thermal photons for different lengths of delay lines.  Solid lines are linear 
fits to the data.}
  \label{FIGURE_ACCIDENTAL}
\end{figure}

We evaluated the performance of the coincidence detector in a quantum optics 
experimental setup that we modified to measure the accidental coincidences from 
two independent Bose-Einstein distributed thermal sources of photons and 
compared the results with theoretical expectations.  Two Si SPCMs were used as 
independent detectors of thermal photons from room light that leaked into the 
single-photon counters.  The detected photocount statistics follow the 
Bose-Einstein distribution for a thermal field with average singles count rates 
of $N_1$ and $N_2$ at the two SPCMs, respectively.  For a coincidence window 
$\tau$ of a few ns, we set the singles detection probabilities to be low such 
that $N_1 \tau, N_2 \tau << 1$, in which case the coincidence probability per 
window is simply given by the product of the singles probabilities, $N_1 N_2 
\tau^2$.  This yields a coincidence rate of $N_1 N_2 \tau$.  In the 
measurements, we counted both the coincidence and singles rates for different 
$N_1$ and $N_2$ and the results are shown in Fig.~\ref{FIGURE_ACCIDENTAL}.
 The horizontal axis is the 
product of the singles rates, $N_1 N_2$, and the plotted coincidence rates show 
a straight line dependence for each coincidence window size with a slope that is 
given by $\tau$.  From the slopes we obtain $\tau$ of 1.13$\pm$0.04 ns, 
2.21$\pm$0.03 ns, and 3.23$\pm$0.12 ns for the 50-cm, 60-cm, and 70-cm delay 
lines, respectively.  The linear dependence of the coincidence counts on $N_1 
N_2$ is in good agreement with theoretical expectations, and the measured $\tau$ 
values agree well with the electronic timing measurements in 
Fig.~\ref{FIGURE_PROBABILITY}.

\section{DISCUSSION}
We have implemented a simple design of a fast electronic coincidence detector 
based on inexpensive PECL components. The coincidence window 
size can be easily adjusted by a change of the two coaxial cable delay lines and 
can be as short as $\sim$1 ns.  The detector has a low propagation delay of 
$\sim$5 ns which allows this circuit to be used as a trigger for a more complex 
coincidence measurement system.  Also, the simplicity and low cost of the 
demonstrated coincidence detection scheme can be easily extended to multiple 
coincidence detection.  Adding timing
overhead did not affect the performance of the coincidence detection.  To 
characterize the coincidence detector independently of the triggering properties 
of the electronic counter, one can add a monostable multivibrator at the output 
of the AND gate, which can be easily implemented using a flip-flop, 
a counter, and an oscillator \cite{ONESHOT}.

\section{ACKNOWLEDGEMENT}

This work was supported by the DoD Multidisciplinary University Research 
Initiative (MURI) program administered by the Army Research Office under Grant 
DAAD-19-00-1-0177 and by ARDA.


\end{document}